\documentclass{article}
\usepackage{cite}
\usepackage{a4wide}
\usepackage{graphicx}
\usepackage{wrapfig}
\usepackage{fancyhdr}
\usepackage{graphicx}
\usepackage[english]{babel}
\usepackage[bookmarks]{hyperref}
\usepackage{graphicx}
\usepackage{bm}
\usepackage[mathlines]{lineno}
\usepackage{subfigure} 
\usepackage{cancel}
\usepackage{amsmath} 
\usepackage{verbatim}
\usepackage{mathtools}
\usepackage{caption}
\usepackage[normalem]{ulem}
\usepackage{amssymb}
\usepackage{amstext} 
\usepackage{amsfonts} 
\usepackage{mathrsfs}
\usepackage{braket}
\usepackage{tikz}
\usepackage{wrapfig}
\usepackage{mwe}
\usepackage{float}

\usepackage[utf8]{inputenc}
\usepackage[left=2.5cm,right=2.5cm,top=2cm,bottom=2cm]{geometry}

\title{\vfill \textbf{On the duality of massive Kalb-Ramond and Proca fields}}
\author{\textbf{Anamaria Hell}\footnote{Hell.Anamaria@physik.uni-muenchen.de}}
\date{\textit{ Ludwig-Maximilians-Universität,\\
Theresienstraße 37, 80333 Munich, Germany}}
\begin{document}

\maketitle
\thispagestyle{empty} 

\begin{abstract}
We compare the massive Kalb-Ramond and Proca fields with a quartic self-interaction and show that the same strong coupling scale is present in both theories. In the Proca theory, the longitudinal mode enters the strongly coupled regime beyond this scale, while the two transverse modes propagate further and survive in the massless limit. In contrast, in  case of the massive Kalb-Ramond field, the two transverse modes become strongly coupled beyond the Vainshtein scale, while the pseudo-scalar mode remains in the weak coupling regime and survives in the massless limit. This indicates a contradiction with the numerous claims in the literature (see eg. \cite{Kawai, Trugenberger, Quevedo, 2001Smailagic, 2002Casini, Auria, Buchbinder, Dalmazi, Shifman, Garcia, Kuzenko, 2002string, Zinoviev, McReynolds, 2017Aurilia, 2019Markou, 2019Schmidt, Heisenberg, 2020SmajlSpal}) that these theories are dual to each other.

\end{abstract}
\vspace*{\fill}

\clearpage
\pagenumbering{arabic} 
\newpage

\section{Introduction}
On the first sight, the massive Kalb-Ramond field theory and the Proca theory may seem completely unrelated. The Kalb-Ramond field is an anti-symmetric two-form \cite{KalbRamond}, whereas the Proca field is a massive vector field \cite{Proca}. Despite this, a number of claims of \textit{duality} between the two theories have recently appeared in the literature (see eg.\cite{Kawai, Trugenberger, Quevedo, 2001Smailagic, 2002Casini, Auria, Buchbinder, Dalmazi, Shifman, Garcia, Kuzenko, 2002string, Zinoviev, McReynolds, 2017Aurilia, 2019Markou, 2019Schmidt, Heisenberg, 2020SmajlSpal}). In other words, both theories are said to describe the same physical system \cite{nonspecialist}. Over the past fifty years, dualities have been shown to hold the promise of greatly improving our understanding of physical models. From the effects of bosonization and magnetic monopoles to the exchange of weak and strong coupling, and the dynamics of topological solutions, the study of dualities has flourished in quantum field theory, string theory, and condensed matter physics (see eg. \cite{Coleman, Dirac, Thooft, Polyakov, MontonenOlive, Witten, Sen, Vafa, Polcinski, Thomas}). 
\\\\
Although Proca and Kalb-Ramond fields are intrinsically different, they both propagate three degrees of freedom. One is a longitudinal mode and two are transversal. Moreover, the actions of these theories are related via the dualization procedure \cite{Kawai, Trugenberger, Quevedo, 2001Smailagic}. Starting with the Kalb-Ramond action, it is possible to construct a parent action where the Kalb-Ramond field as well as its field strength are each independent fields, and with which one can restore the equations of motion of the original theory. Integrating out the Kalb-Ramond field and expressing its field strength in terms of its dual, one obtains the Proca theory. Using a similar approach, it is possible to return from Proca to Kalb-Ramond theory. Given these considerations, it might be tempting to argue that both theories describe the same physics.
However, this is not the case for their corresponding massless theories.  
While the massless Kalb-Ramond field includes only one degree of freedom, the longitudinal mode, the Maxwell theory has two degrees of freedom, the transverse modes. 
Furthermore, the dualization procedure which would connect the actions of these two theories doesn't exist. Rather, while the massless antisymmetric two-form is dual to a scalar field, the Maxwell field is self-dual \cite{KalbRamond, Kawai, DavisShellard}. To show this, one would start again from the parent action which now contains a kinetic term and a Bianchi identity implemented by the Lagrangian multiplier, as opposed to a massive case. As we integrate out the field strength tensor of the theory which we dualize,
the resulting action only becomes a functional of the Lagrange multiplier. This is the dual theory, in which the Lagrange multiplier takes the role of the dual field \cite{nonspecialist}. 
The contrast among the massless theories raises the question about the behavior of the massive theories in the massless limit. Will the transverse and longitudinal modes behave in the same way? In order to gain an insight, we will alter the theories by a self-interaction and compare the behavior of the degrees of freedom.  We will notice that these theories share similar properties with massive gravity and massive Yang-Mills theories. 
\\\\
Both linearized massive gravity with the mass term of the Fiersz-Pauli form \cite{FierszPauli}, and massive Yang-Mills theory with a mass added "by hand", suffer from a very peculiar property. If we perturbatively evaluate these theories, and take the massless limit, we would find a deviation with respect to the prediction of the respective massless theories \cite{vDVZ, Zakharov, Iwasaki, Wong, SlavFad}. This is known as the vDVZ discontinuity.  The reason for the discrepancy in both theories is the longitudinal degree of freedom, which is absent in the respective massless theories. 
In the massless limit, it does not decouple from the remaining degrees of freedom. Thus, it makes an additional contribution to the experimental predictions of the theories that is not available in the massless case. But, we shouldn't ignore the fact that these theories are non-linear. According to Vainshtein, there exists a scale, known as the strong coupling scale (or equivalently, the Vainshtein scale), where the nonlinear terms become important. On the level of equations of motion, it is the scale at which the nonlinear terms become of the same order as the linear terms. As soon as this scale is reached, perturbation theory breaks down, and the field enters a strongly coupled regime. By considering these nonlinear terms, Vainshtein has shown that for both theories the longitudinal mode enters the strong coupling regime, where it remains for scales smaller than the Vainshtein radius \cite{Vainshtein, VainshteinMeh, PertNep}. As a consequence, it decouples away from the rest of the matter up to a minor correction which disappears in the massless limit\cite{Gruzinov, Hell}. When the mass is set to zero, the Vainshtein radius turns infinite, and so both general relativity and massless Yang-Mills theory are recovered. 
\\\\
The purpose of this paper is to study the way in which the Vainshtein mechanism applies to the Kalb-Ramond and Proca theories. This comparison is motivated by the assertion of duality of the two theories, which is frequently found in the literature. By adding a quartic self-interaction to each of the theories, we shall find that the two theories suffer from a discontinuity in the massless limit. In the case of Proca theory, the source of this discontinuity is the longitudinal mode, which is absent in the Maxwell theory. This mode is similar to the one of the massive Yang-Mills theory. The massive Kalb-Ramond theory also includes a discontinuity in the massless limit. However, the source of that discontinuity are the transverse modes. In the example of mimetic massive gravity, it was demonstrated how the strong coupling scale can be evaluated by the minimum quantum fluctuations of the fields \cite{Mimetic}. Such fluctuations are a direct consequence of Heisenberg's uncertainty principle, which we can evaluate through the equal-time two point correlation function \cite{QFTCS}. With their help, we will find that the same Vainshtein scale is shared by both theories. However, the distinction between them still remains as there are different modes entering the strong coupling regime. For the Kalb-Ramond field, these are the transverse modes. Therefore, for a non-zero mass, this theory will propagate just the longitudinal modes beneath the Vainshtein scale. In contrast, the Proca theory will propagate only transverse modes since longitudinal modes will enter the strong coupling regime. Although these theories are massive, they differ in their characteristics. We will see that this is indicated  by the minimal level of quantum fluctuations at the level of free theory. This casts a doubt on the existence of the duality between the two theories.
\\\\
The duality claim was also extended to Stueckelberg theory, massive $p$-forms \cite{Lahiri, Hari, 1999Smag} and curved spacetime \cite{Buchbinder}. In particular, in the case of Stueckelberg theory for a $p=3$ form, the theory was considered with an arbitrary potential \cite{Dvali}. Similarly to the case we have considered in this paper, it is reasonable to suspect that a similar confusion might occur for massive $p$ forms. They are said to be dual $D-p-1$ forms, whereas massless ones are dual to $D-p-2$ forms \cite{Kawai, nonspecialist, Ansoldi}. In the Appendix, we will demonstrate on how the analysis presented in this work applies to a massive three-form.

\section{Comparison of the free theories}
In this chapter we will study and compare the free theories of massive Kalb-Ramond and Proca fields. The action for the massive Kalb-Ramond theory is given by 
\begin{equation}\label{eq::KBaction}
    S=\frac{1}{12}\int d^4x\left(H_{\mu\nu\rho}H^{\mu\nu\rho}-3m^2B_{\mu\nu}B^{\mu\nu}\right),
\end{equation}
where 
\begin{equation}
    H_{\mu\nu\rho}=B_{\nu\rho,\mu}+B_{\rho\mu,\nu}+B_{\mu\nu,\rho}
\end{equation}
is the field strength tensor for the Kalb-Ramond field $B_{\mu\nu}$, which is anti-symmetric in its indices. We will use commas to denote the derivative with respect to the corresponding coordinate $x^{\mu}$ throughout the paper. This theory will be compared with the Proca theory, whose action is given by 
\begin{equation}\label{eq::Paction}
    S=\int d^4x\left(-\frac{1}{4}F_{\mu\nu}F^{\mu\nu}+\frac{m^2}{2}A_{\mu}A^{\mu}\right),
\end{equation}
where 
\begin{equation}
    F_{\mu\nu}=A_{\nu,\mu}-A_{\mu,\nu}
\end{equation}
is the field strength tensor for the vector field $A_{\mu}$. Now will we analyse the content of these theories. Following \cite{HermGrav} we shall decompose the spatial and time components of the Kalb-Ramond field as follows.
\begin{equation}
    \begin{split}
        &B_{0i}=C_i^T+\mu_{,i},\qquad\text{}\qquad C_{i,i}^T=0\\\\
        &B_{ij}=\varepsilon_{ijk}B_k,\qquad\qquad B_i=B_i^T+\phi_{,i},\qquad\text{}\qquad B_{i,i}^T=0
    \end{split}
\end{equation}
where $i,j,k={1,2,3}$. We will refer to the pseudovector $B_i^T$ as the transverse mode, and a pseudoscalar $\phi$ as the longitudinal mode.  Substituting this decomposition into (\ref{eq::KBaction}), we obtain
\begin{equation}\label{eq::KBaction2}
\begin{split}
     S=\frac{1}{2}\int d^4x&\left[C_i^T(-\Delta+m^2)C_i^T-2\varepsilon_{ijk}C_i^T\dot{B}^T_{k,j}-m^2\mu\Delta\mu\right.\\
     &\left.+\dot{B}_i^T\dot{B}_i^T-m^2B_i^TB_i^T-\dot{\phi}\Delta\dot{\phi}-\Delta\phi\Delta\phi+m^2\phi\Delta\phi\right].
\end{split}
\end{equation}
Here, the point corresponds to the time derivative. We can see that $C_i^T$ and $\mu$ do not propagate because there are no time derivatives acting on them. As we are only concerned with the physical degrees of freedom, we will integrate them out as follows. If we vary the action with respect to $C_i^T$ and $\mu$, we find that they satisfy the following constraints, respectively:
\begin{equation}
    (-\Delta+m^2)C_i^T=\varepsilon_{ijk}\dot{B}^T_{k,j}\qquad\text{and}\qquad \Delta\mu=0
\end{equation}
whose solution is given by
\begin{equation}\label{eq::KBconstlinear}
    C_i^T=\frac{\varepsilon_{ijk}}{-\Delta+m^2}\dot{B}^T_{k,j}\qquad\text{and}\qquad \mu=0.
\end{equation}
The operator acting on the transverse modes in the first solution should be understood in terms of Fourier modes. If we would express it as 
\begin{equation}
    B_i^T=\int d^3k \Tilde{B}_i^T(\Vec{k})e^{i\Vec{k}\Vec{x}}
\end{equation}
this operator would act as 
\begin{equation}
    \frac{1}{-\Delta+m^2}B_i^T=\int d^3k\frac{1}{|\Vec{k}|^2+m^2}\Tilde{B}_{ i}^T(\Vec{k})e^{i\Vec{k}\Vec{x}}.
\end{equation}
Substituting these solutions into (\ref{eq::KBaction2}) results in the following action:
\begin{equation}\label{eq::KBaction3}
\begin{split}
     S=-\frac{1}{2}\int d^4x&\left[B_i^T(\Box+m^2)\frac{m^2}{-\Delta+m^2}B_i^T+\phi(\Box+m^2)(-\Delta\phi)\right]
\end{split}
\end{equation}
This action only contains propagating fields. 
We can notice that the fields are not canonically normalized. In terms of the canonically normalized variables
\begin{equation}\label{eq::cannorm}
    B_{ni}^T=\sqrt{\frac{m^2}{-\Delta+m^2}}B_i^T \qquad\text{and}\qquad \phi_n=\sqrt{-\Delta}\phi
\end{equation}
 the action is given by 
\begin{equation}\label{eq::KBaction4}
\begin{split}
     S=-\frac{1}{2}\int d^4x&\left[B_{ni}^T(\Box+m^2)B_{ni}^T+\phi_n(\Box+m^2)\phi_n\right].
\end{split}
\end{equation}
The operators in (\ref{eq::cannorm}) are again to be understood in terms of Fourier modes with wave number $\Vec{k}$.
To see how this field changes with a length scale $L\sim\frac{1}{k}$, where $k$ is the magnitude of the wavenumber, we will now consider what the minimum fluctuations for the field are. 
In general, for a normalised quantum field, the typical amplitude of the quantum fluctuations is of order $\left(\frac{k^3}{\omega_k}\right)^{\frac{1}{2}}$, for scales $k$ corresponding to length scales $L$ as $k\sim\frac{1}{L}$ \cite{QFTCS}. Therefore, the minimal level of quantum fluctuations at scales $\frac{1}{L^2}\gg m^2$ for normalised transverse and longitudinal modes is given respectively by 
\begin{equation}
\begin{split}
    \delta B^T_{nL}\sim\frac{1}{L} \qquad\text{and}\qquad \delta\phi_{nL}\sim\frac{1}{L}
\end{split}
\end{equation}
Using (\ref{eq::cannorm},) we then find for the original transverse mode:
\begin{equation}
    \delta B^T_{L}\sim\frac{1}{mL^2}
\end{equation}
As already pointed out in \cite{HermGrav}, we can see that the amplitude of the quantum fluctuations for the transverse modes grows faster than the amplitude for the longitudinal modes in case of a small mass. This implies that the transverse modes enter the strong coupling regime ahead of the longitudinal modes, once one considers an interacting theory.
\\\\
We shall now consider the free Proca theory. We will decompose the spatial part of the vector field $A_{\mu}$ as
\begin{equation}\label{eq::Pdecomp}
    A_i=A_i^T+\chi_{,i},\qquad\text{with}\qquad A_{i,i}^T=0.
\end{equation}
The transverse modes are now given by $A_i^T$ and the longitudinal ones are given by the field $\chi$.
If we substitute this decomposition into (\ref{eq::Paction}), we obtain
\begin{equation}\label{eq::Paction01}
    \begin{split}
        S=\frac{1}{2}\int d^4x &\left[A_0\left(-\Delta+m^2\right)A_0+2A_0\Delta\dot{\chi}-\left(\dot{\chi}\Delta\dot{\chi}-m^2\chi\Delta\chi\right)\right.\\ &\left.+\left(\dot{A_i^T}\dot{A_i^T}-A_{i,j}^TA_{i,j}^T-m^2A_i^TA_i^T\right)\right].
        \end{split}
\end{equation}
Compared to the Kalb-Ramond field, which had two non-propagating fields, Proca theory has only one component that does not propagate. It is the $A_0$ component. Using the same procedure as before, we now integrate it out.  
Varying (\ref{eq::Paction01}) with respect to it, we find the following constraint: 
\begin{equation}
     (-\Delta+m^2)A^0=-\Delta\dot{\chi}
\end{equation}
Its solution is given by
\begin{equation}
    A_0=\frac{-\Delta}{-\Delta+m^2}\dot{\chi}.
\end{equation}
Plugging it back into the action, we obtain the action consisting only of the transverse and longitudinal modes, 
\begin{equation}\label{eq::Paction2}
    S=-\frac{1}{2}\int d^4x\left[A_i^T(\Box+m^2)A_i^T+\chi(\Box+m^2)\frac{m^2(-\Delta)}{-\Delta+m^2}\chi\right].
\end{equation}
 We can notice that the kinetic term of the longitudinal mode is not canonically normalised. Defining the normalised variable as
\begin{equation}\label{eq::normlong}
    \chi_n=m\sqrt{\frac{-\Delta}{-\Delta+m^2}}\chi,
\end{equation}
 we can express (\ref{eq::Paction2}) as
\begin{equation}\label{eq::Paction3}
       S=-\frac{1}{2}\int d^4x\left[A_i^T(\Box+m^2)A_i^T+\chi_n(\Box+m^2)\chi_n\right].
\end{equation}
Let us see how the original fields evolve in terms of mass and scale. The fields in (\ref{eq::Paction2}) are canonically normalized.  Therefore, that the minimal level of quantum fluctuations at scales $\frac{1}{L^2}\gg m^2$ is given for them by 
\begin{equation}
     \delta A^T_{nL}\sim\frac{1}{L} \qquad\text{and}\qquad \delta\chi_{nL}\sim\frac{1}{L}.
\end{equation}
Using (\ref{eq::normlong}), we obtain that the minimum of the quantum fluctuations for the original longitudinal mode is
\begin{equation}
    \delta \chi_{L}\sim\frac{1}{mL}.
\end{equation}
If we divide it by the length scale $L$ in order to compare it with the transverse modes, we can see that it grows faster than these when the mass is small.
\\\\
Up to now we saw that both theories propagate the same number of degrees of freedom. Nevertheless, the difference between these theories lies in the minimum quantum fluctuations for the original modes. 
The quantum fluctuations associated with the longitudinal mode of the Proca theory depend on mass in the same way as the transverse modes of the Kalb-Ramond field. Then again, the transverse modes of Proca theory and the longitudinal modes of the Kalb-Ramond field both grow in the same way when the length scale decreases.  
We shall see in the next chapter how this difference among the same kinds of modes leads to different properties of the theories as soon as we modify them by an interaction.

\section{Above and beyond the strong coupling scale}

Now, we will modify both of these theories by adding a quartic self-interaction. For the case of the Kalb-Ramond field, we will consider the action
\begin{equation}\label{eq::KBactionInt}
    S=\int d^4x\left[\frac{1}{12}H_{\mu\nu\rho}H^{\mu\nu\rho}+\frac{m^2}{4}B_{\mu\nu}B^{\mu\nu}+\frac{g^2}{16}\left(B_{\mu\nu}B^{\mu\nu}\right)^2\right],
\end{equation}
which we will compare with the interacting Proca theory
\begin{equation}\label{eq::PactionInt}
    S=\int d^4x\left[-\frac{1}{4}F_{\mu\nu}F^{\mu\nu}+\frac{m^2}{2}A_{\mu}A^{\mu}+\frac{g^2}{4}\left(A_{\mu}A^{\mu}\right)^2\right].
\end{equation}
 We will assume that the coupling constant satisfies $g^2\ll1$, and will consider these theories on energy scales $k^2\gg m^2$. First we will perturbatively analyse the theories. We will demonstrate how both theories suffer from a discontinuity in the massless limit. With the help of the quantum fluctuations which we have found in the previous chapter, we will determine the Vainshtein radius. This will be followed by an analysis beyond perturbation theory, where we will study the corrections to the modes appearing in the corresponding massless theories. 

\subsection{Proca field}
As a first step, we will reformulate the Lagrangian in terms of physical modes, proceeding as in the free theory. Since there is no propagation of the $A_0$ component, let us integrate it out. Varying the action (\ref{eq::PactionInt}) with respect to it, we find a constraint satisfied by $A_0$ 
\begin{equation}\label{eq::Pconstraint}
    (-\Delta+m^2+g^2A_iA_i)A_0-g^2A_0^3=-\Delta\dot{\chi},
\end{equation}
where $A_i$ is given by (\ref{eq::Pdecomp}). Under the assumption that 
\begin{equation}\label{eq::assumptionP}
    g^2\chi^2<1,
\end{equation}
which will be verified aposteriori, we can resolve this constraint for scales $k^2\sim\frac{1}{L^2}\gg m^2$ as
\begin{equation}
\begin{split}
    A_0&=\frac{-\Delta}{-\Delta+m^2}\dot{\chi}+\frac{g^2}{-\Delta+m^2}\left[\left(\dot{\chi}^2-\chi_{,i}\chi_{,i}-2\chi_{,i}A_i^T-A_i^TA_i^T\right)\dot{\chi}\right]\\\\
    &+\mathcal{O}\left(g^2(mL)^2\frac{\chi^3}{L},g^4\frac{\chi^5}{L}\right).
\end{split}
\end{equation}
Here $\frac{1}{L}$ denotes a derivative acting on $\chi$. Substituting this expression back into the action, we arrive at the Lagrangian density 
\begin{equation}
    \mathcal{L}=\mathcal{L}_0+\mathcal{L}_{int},  \qquad\text{where}
\end{equation}
\begin{equation*}
    \begin{split}
        \mathcal{L}_0&=-\frac{1}{2}\chi(\Box+m^2)\frac{m^2(-\Delta)}{-\Delta+m^2}\chi-\frac{1}{2}A_i^T(\Box+m^2)A_i^T \qquad\text{and}\\\\
        \mathcal{L}_{int}&=\frac{g^2}{4}\left(\chi_{,\mu}\chi^{,\mu}\right)^2-g^2\chi_{,\mu}\chi^{,\mu}\chi_{,i}A_i^T-\frac{g^2}{2}\chi_{,\mu}\chi^{,\mu}A_i^TA_i^T+g^2\left(\chi_{,i}A_i^T\right)^2\\\\
        &+\mathcal{O}\left(g^2m^2\frac{\chi^4}{L^2},g^2\frac{\chi}{L}\left(A^T\right)^3\right).
    \end{split}
\end{equation*}
In $\mathcal{L}_{int}$ we have only kept the most relevant terms. It is clear from the kinetic term that the longitudinal mode needs to be properly normalised. Using the normalised variables $\chi_n$ defined in (\ref{eq::normlong}), we obtain for scales $\frac{1}{L^2}\gg m^2$
\begin{equation}
    \mathcal{L}=\mathcal{L}_0+\mathcal{L}_{int}, \qquad\text{where now}
\end{equation}
\begin{equation*}
    \begin{split}
        \mathcal{L}_0&=-\frac{1}{2}\chi_n(\Box+m^2)\chi_n-\frac{1}{2}A_i^T(\Box+m^2)A_i^T\qquad\text{and}\\\\
        \mathcal{L}_{int}&\sim\frac{g^2}{4m^4}\left(\chi_{n,\mu}\chi_n^{,\mu}\right)^2-\frac{g^2}{m^3}\chi_{n,\mu}\chi_n^{,\mu}\chi_{n,i}A_i^T-\frac{g^2}{2m^2}\chi_{n,\mu}\chi_n^{,\mu}A_i^TA_i^T+\frac{g^2}{m^2}\left(\chi_{n,i}A_i^T\right)^2.
    \end{split}
\end{equation*}
The interacting terms indicate that there exists a discontinuity in the massless limit. However, we should not forget that this theory has so far been treated in a perturbative way. In order to find the scale at which the interacting terms become relevant,
 we first need to identify the most dominant term among them. This will be done with the help of the minimal level of quantum fluctuations, which we have found in the previous chapter. Then, we determine the Vainshtein radius by finding the scale for which the most dominant term is of the same order as the kinetic term. If we estimate the derivatives as $\partial_{\mu}\sim\frac{1}{L}$ and the normalised longitudinal and transverse modes as $\chi_n\sim\frac{1}{L}$ and $A_i^T\sim\frac{1}{L}$, respectively, we can evaluate the interaction terms as follows:
\begin{equation}\label{eq::intProca}
\begin{split}
     \frac{g^2}{4m^4}\left(\chi_{n,\mu}\chi_n^{,\mu}\right)^2&\sim\frac{g^2}{(mL)^4L^4},\qquad\qquad \frac{g^2}{m^3}\chi_{n,\mu}\chi_n^{,\mu}\chi_{n,i}A_i^T\sim\frac{g^2}{(mL)^3L^4}\qquad\text{and}\\\\
     \frac{g^2}{2m^2}\chi_{n,\mu}\chi_n^{,\mu}A_i^TA_i^T&\sim\frac{g^2}{m^2}\left(\chi_{n,i}A_i^T\right)^2\sim\frac{g^2}{(mL)^2L^4}.
\end{split}
\end{equation}
Evidently, for a small mass, the first term will be the most dominant. It becomes of the same order of magnitude as the kinetic term for the longitudinal modes at length scales 
\begin{equation}
    L_{str}\sim\frac{\sqrt{g}}{m}.
\end{equation}
At this scale, the longitudinal modes enters the strong coupling regime. This scale agrees with that of \cite{Khoury}, where it was obtained by introducing a Stueckelberg field and identifying the interacting term with the largest singularity in the massless limit. 
On the other hand, the most significant contributor to the transverse modes is the second term. It is of the same order of magnitude as the kinetic term of the transverse modes at scales 
\begin{equation}
    L^T\sim\frac{g^{2/3}}{m}.
\end{equation}
Because this scale is smaller than the strong coupling scale of the longitudinal modes, we cannot claim that the transverse modes enter the strong coupling regime as the perturbation theory for the longitudinal modes is no longer reliable. Therefore, to find out what happens to the transverse modes, we must analyse the theory beyond the strong coupling scale. However, before we do that, we will verify the strong coupling scale on the level of the equations of motion which define it. 
\\\\
The equations of motion for the longitudinal and transverse modes on scales $\frac{1}{L^2}\gg m^2$ with the most important terms are given by 
\begin{equation}
\begin{split}
    (\Box+m^2)\chi\sim&-\frac{g^2}{m^2}\left(\chi^{,\mu}\chi_{,\mu}\Box\chi+2\chi_{,\mu}\chi_{,\nu}\chi^{,\mu\nu}\right)\\&+\frac{2g^2}{m^2}\left(\chi_{,i}A_i^T\Box\chi+2\chi_{,i\mu}\chi^{,\mu}A_i^T+\chi^{,\mu}\chi_{,i}A_{i,\mu}^T\right)
\end{split}
\end{equation}
and
\begin{equation}
    (\Box+m^2)A_k^T\sim-g^2\left(\delta_{ki}-\frac{\partial_k\partial_i}{\Delta}\right)\left(\chi_{,i}\chi_{,\mu}\chi^{\mu}+A_i^T\chi_{,\mu}\chi^{,\mu}-2\chi_{,i}\chi_{,j}A_j^T\right).
\end{equation}
To find the Vainhstein radius, we will develop the perturbation theory as we expand the longitudinal and transverse modes in powers of $g^2$
\begin{equation}
    \begin{split}
        \chi&=\chi^{(0)}+\chi^{(1)}+...,\\
        A_i^T&=A_i^{T(0)}+A_i^{T(1)}+...,
    \end{split}
\end{equation}
where $\chi^{(0)}$ and $A_i^{T(0)}$ satisfy the linear equation of motion
\begin{equation}
    (\Box+m^2)\chi^{(0)}=0\qquad\qquad\text{and}\qquad (\Box+m^2)A_i^{T(0)}=0,
\end{equation}
the solution of which are plane waves. For simplicity, we will omit the exponent $(0)$ associated with the free fields. The corrections to the longitudinal modes satisfy
\begin{equation}
\begin{split}
    (\Box+m^2)\chi^{(1)}\sim&-\frac{2g^2}{m^2}\chi_{,\mu}\chi_{,\nu}\chi^{,\mu\nu}+\frac{2g^2}{m^2}\left(2\chi_{,i\mu}\chi^{,\mu}A_i^T+\chi^{,\mu}\chi_{,i}A_{i,\mu}^T\right).
\end{split}
\end{equation}
By estimating $\chi\sim\frac{1}{mL}$ and $A_i^T\sim\frac{1}{L}$, we find that the first nonlinear term is more dominant than the second one. Thus, we can calculate the corrections to the longitudinal modes as 
\begin{equation}
    \chi^{(1)}\sim\frac{g^2\chi^3}{(mL)^2}\sim\frac{g^2}{(mL)^5}.
\end{equation}
The perturbation theory in the longitudinal mode breaks down when the nonlinear term $\chi^{(1)}$ reaches the order of magnitude of the linear term. This happens at the Vainshtein radius $L_{str}\sim\frac{\sqrt{g}}{m}$, as can be observed already from the Lagrangian. 
We can estimate the corrections to the transverse modes from the longitudinal modes as follows:
\begin{equation}
    A_k^{T(1)}\sim\frac{g^2\chi^3}{L}\sim\frac{g^2}{(mL)^3}
\end{equation}
Even though these corrections seem to be troublesome at very high energies, leading to divergences in the massless limit, one cannot trust them. This is because they become of the same order of magnitude as the linear term at scales lower than the strong coupling scale for the longitudinal mode. \\
Lastly, we can see that the assumption (\ref{eq::assumptionP}) is violated at scales $L\sim\frac{g}{m}$ which are also smaller then the strong coupling scale. As a result, the expansion of the $A_0$ constraint can be carried out safely below the strong coupling scale. 
\\\\
As soon as the longitudinal modes enter the strong coupling regime, the most dominant term for the longitudinal modes in the Lagrangian will be 
\begin{equation}
    \mathcal{L}_{\chi int}=\frac{g^2}{4}\left(\chi_{,\mu}\chi^{,\mu}\right)^2.
\end{equation}
This term specifies the new canonical normalization, where the canonically normalized variable is given by
\begin{equation}
    \chi_{n}\sim\frac{g}{\sqrt{2}}\frac{\chi^2}{L}.
\end{equation}
 Hence, the most dominant term can be written as 
\begin{equation}
     \mathcal{L}_{\chi int}\sim\frac{1}{2}\chi_{n,\mu}\chi_n^{,\mu}.
\end{equation}
The minimal level of quantum fluctuations for the normalised mode is $\delta\chi_n\sim\frac{1}{L}$ and therefore, for the original mode they become
\begin{equation}
    \delta_L\chi\sim\frac{1}{\sqrt{g}}.
\end{equation}
Since $g<1$, it is easy to see that the assumption (\ref{eq::assumptionP}) will always be satisfied. \\
Even if the perturbation theory for longitudinal modes fails for scales $L\leq L_{str}$, it is still possible to use it for the transverse modes. The most important terms for the transversal modes are given by 
\begin{equation}
    \begin{split}
        &\mathcal{L}_{A^T}\sim-\frac{1}{2}A_i^T(\Box+m^2)A_i^T-g^2\chi_{,\mu}\chi^{,\mu}\chi_{,i}A_i^T+\mathcal{O}\left(g^2\frac{\chi^2}{L^2}\left(A^T\right)^2\right).
    \end{split}
\end{equation}
If we now estimate the longitudinal mode as $\chi\sim\frac{1}{\sqrt{g}}$, we can see that the interaction term, which was troublesome in the massless limit for scales $L>L_{str}$, will always be smaller than the kinetic term:
\begin{equation}
    g^2\chi_{,\mu}\chi^{,\mu}\chi_{,i}A_i^T\sim\frac{\sqrt{g}}{L^4}
\end{equation}
In other words, as soon as the longitudinal mode enters the strong coupling regime, the transverse mode corrections due to the longitudinal mode scale as 
\begin{equation}
    A_i^{T(1)}\sim\frac{\sqrt{g}}{L}.
\end{equation}
We have thus seen that due to the strong coupling of the longitudinal mode, any divergences that occur in the massless limit disappear. Below the strong coupling scale, only the two transverse modes continue to propagate. The Vainshtein mechanism has also been considered in the context of generalised Proca theory with derivative self-interactions, where it led to the suppression of the longitudinal mode \cite{GenPV}.  

\subsection{Kalb-Ramond field}
Following the approach similar to the one used in the first part of this chapter, we will now study the theory of the self-interacting Kalb-Ramond field. First, we will express the Lagrangian only in terms of the propagating degrees of freedom. Decomposing the Kalb-Ramond field as in the free case, we find that this system has two constraints. Varying the action (\ref{eq::KBactionInt}) with respect to $C_i^T$, we obtain 
\begin{equation}\label{eq::Cconst}
    \begin{split}
        (-\Delta+m^2)C_l^T&=\varepsilon_{ljk}\dot{B}^T_{k,j}-g^2P_{li}^T\left[C_i^TC_j^TC_j^T+2C_i^TC_j^T\mu_{,j}+C_j^TC_j^T\mu_{,i}+2C_j^T\mu_{,i}\mu_{,j}\right.\\\\
        &\left.+C_i^T\mu_{,j}\mu_{,j}-(C_i^T+\mu_{,i})\left(B_i^TB_i^T+2B_i^T\phi_{,i}+\phi_{,i}\phi_{,i}\right)+\mu_{,i}\mu_{,j}\mu_{,j}\right],
    \end{split}
\end{equation}
where 
\begin{equation}
    P_{ij}^T=\delta_{ij}-\frac{\partial_i\partial_j}{\Delta}
\end{equation}
is the transverse projector. The constraint satisfied by $\mu$ is given by 
\begin{equation}\label{eq::muconst}
    \begin{split}
        -m^2\Delta\mu&=g^2\partial_i\left[\mu_{,i}\mu_{,j}\mu_{,j}+2\mu_{,i}\mu_{,j}C_j^T+\mu_{,j}\mu_{,j}C_i^T+\mu_{,i}C_j^TC_j^T+2\mu_{,j}C_j^TC_i^T\right.\\\\
        &\left.-(\mu_{,i}+C_i^T)\left(B_j^TB_j^T+2B_j^T\phi_{,j}+\phi_{,j}\phi_{,j}\right)+C_i^TC_j^TC_j^T\right].
    \end{split}
\end{equation}
Up to $\mathcal{O}(g^4)$, the first constraint can be resolved as
\begin{equation}
    \begin{split}
        C_i^T&=\varepsilon_{ijk}D\left(\dot{B}_{k,j}^T\right)\\\\
        &+\frac{g^2\varepsilon_{ljk}P^T_{il}}{-\Delta+m^2}\left\{D\left(\dot{B}^T_{k,j}\right)\left[\left(B_s^T+\phi_{,s}\right)^2-\varepsilon_{sab}\varepsilon_{scd}D\left(\dot{B}_{b,a}^T\right)D\left(\dot{B}_{d,c}^T\right)\right]\right\},
    \end{split}
\end{equation}
where
\begin{equation}\label{eq::Dop}
    D(B_i)=\frac{1}{-\Delta+m^2}\left[B_i\right].
\end{equation}
The solution of the constraint satisfied by $\mu$ up to $\mathcal{O}(g^4)$ is given by 
\begin{equation}
    \mu=\frac{g^2\varepsilon_{ijk}}{m^2\Delta}\partial_i\left\{D\left(\dot{B}^T_{k,j}\right)\left[B_s^TB_s^T+2B_s^T\phi_{,s}+\phi_{,s}\phi_{,s}-\varepsilon_{sab}\varepsilon_{scd}D\left(\dot{B}_{b,a}^T\right)D\left(\dot{B}_{d,c}^T\right)\right]\right\}.
\end{equation}
Substituting these into the action and keeping only the most dominant interactions we obtain the Lagrangian density 
\begin{equation}\label{eq::KBLang}
    \mathcal{L}=\mathcal{L}_0+\mathcal{L}_{int},\qquad\text{where}
\end{equation}
\begin{equation*}
    \begin{split}
        \mathcal{L}_0&=-\frac{1}{2}B_i^T(\Box+m^2)\frac{m^2}{-\Delta+m^2}B_i^T-\frac{1}{2}\phi(\Box+m^2)(-\Delta\phi)\qquad\text{and}\\\\
        \mathcal{L}_{int}&=\frac{g^2}{4}\left\{\left[\varepsilon_{ijk}\varepsilon_{ils}D\left(\dot{B}^T_{k,j}\right)D\left(\dot{B}^T_{s,l}\right)\right]^2-2\varepsilon_{ijk}\varepsilon_{ils}D\left(\dot{B}^T_{k,j}\right)D\left(\dot{B}^T_{s,l}\right)B_p^TB_p^T+\left(B_i^TB_i^T\right)^2\right\}\\\\
        &-g^2\left[\varepsilon_{ijk}\varepsilon_{ils}D\left(\dot{B}^T_{k,j}\right)D\left(\dot{B}^T_{s,l}\right)B_p^T\phi_{,p}-B_i^TB_i^TB_j^T\phi_{,j}\right]\\\\
        &-\frac{g^2}{2}\left[\varepsilon_{ijk}\varepsilon_{ils}D\left(\dot{B}^T_{k,j}\right)D\left(\dot{B}^T_{s,l}\right)\phi_{,p}\phi_{,p}-B
        _i^TB_i^T\phi_{,j}\phi_{,j}-2\left(B_i^T\phi_{,i}\right)^2\right]+\mathcal{O}\left(\frac{g^2B^T\phi^3}{L^3}\right).
    \end{split}
\end{equation*}
As in the free theory, the longitudinal and transverse modes should be canonically normalised according to (\ref{eq::cannorm}). This determines the minimum level of quantum fluctuations for the original modes. We can use either the original or the normalised modes for finding the strong coupling scale, and the result will be the same irrespective of the choice. Since the normalised modes make the terms appear more complicated, we will present the search for the Vainshtein radius using the original fields. 
 By considering the scales $k^2\sim\frac{1}{L^2}\gg m^2$, we can estimate the derivatives as $\partial_{\mu}\sim\frac{1}{L}$ and the operator (\ref{eq::Dop}) as $D\sim L^2$. Then, the terms in each row of $\mathcal{L}_{int}$ are proportional to one another. Representative terms of each row can be evaluated as
\begin{equation}
\begin{split}
    \frac{g^2}{4}\left(B_i^TB_i^T\right)^2&\sim g^2\left(B^T\right)^4,\qquad g^2B_i^TB_i^TB_j^T\phi_{,j}\sim\frac{g^2\left(B^T\right)^3\phi}{L},\qquad g^2\left(B_i^T\phi_{,i}\right)^2\sim\frac{g^2\left(B^T\right)^2\phi^2}{L^2}.
\end{split}
\end{equation}
Keeping in mind that the quantum fluctuations of the original transverse and longitudinal modes are given by $\delta B_L^T\sim\frac{1}{mL^2}$ and $\delta\phi_L\sim\mathcal{O}\left(1\right)$ on scales $k^2\sim\frac{1}{L^2}\gg m^2$, respectively, we can estimate these terms as 
\begin{equation}
\begin{split}
    g^2\left(B^T\right)^4&\sim\frac{g^2}{\left(mL\right)^4L^4},\qquad \frac{g^2\left(B^T\right)^3\phi}{L}\sim\frac{g^2}{\left(mL\right)^3L^4},\qquad \frac{g^2\left(B^T\right)^2\phi^2}{L^2}\sim\frac{g^2}{\left(mL\right)^2L^4}.
\end{split}
\end{equation}
We can already notice the first difference to the expressions of the Proca theory (\ref{eq::intProca}). As more longitudinal modes were involved in the interaction, the divergence in the limit $m\to0$ was larger. Here the situation is reversed. The greater divergence in the mass corresponds to the terms involving more transverse modes. 
Of the above terms, the first one is clearly the most dominant for a small mass. At the level of the equation of motion for transverse modes, this would be equivalent to a cubic term. By comparing it with the linear term, or equivalently, by comparing the interacting term with the kinetic term for the transverse modes in the Lagrangian, we find that the  strong coupling scale is given by
\begin{equation}
    L_{str}\sim\frac{\sqrt{g}}{m}.
\end{equation}
 Among the nonlinear terms, the second one contributes most to the longitudinal mode. This term becomes of the same order as the kinetic term at length-scale 
\begin{equation}
    L_{\phi}\sim\frac{g^{2/3}}{m}.
\end{equation}
However, this scale is not to be trusted because perturbation theory is not valid in transverse modes below $L_{str}$. 
As a first step in analyzing the theory beyond the strong coupling scale, we will consider the theory at scales $L\sim L_{str}$. It turns out that at these scales the higher order terms, suppressed for larger scales, become as dominant as the kinetic term. These terms have the form
\begin{equation}
    \mathcal{L}_{int}\supset\sum_{n=2}^{\infty}\frac{g^{2n+2}}{m^{2n}}\left(B^T\right)^{2n+4}\sim\sum_{n=2}^{\infty}\left(\frac{L_{str}}{L}\right)^{4n+4}\frac{1}{L^4}
\end{equation}
We shall see that the reason for their occurrence is the $\mu$-constraint, which can not be expanded for scales $L\leq L_{str}$. Thus, to analyse the theory beyond the scale of strong coupling, we must resort to the (\ref{eq::Cconst}) and (\ref{eq::muconst}) constraints. At the strong coupling scale, it is possible to estimate 
\begin{equation}\label{eq::Cmu}
    C^T\sim B^T\sim\frac{1}{\sqrt{g}L_{str}},\qquad\qquad \mu\sim\frac{g^2L_{str}}{m^2}\left(B^T\right)^3\sim\frac{1}{\sqrt{g}}.
\end{equation}
On scales $\frac{1}{L^2}\gg m^2$, we can estimate the constraint (\ref{eq::Cconst}) as\footnote{We will ignore the coefficients for each of these terms, since all that matters is how these terms scale with $g,m$ and $L$.} 
\begin{equation}
\begin{split}
    \frac{C^T}{L_{str}^2}&\sim\frac{B^T}{L_{str}^2}  -g^2\left[\left(C^T\right)^3+\left(C^T\right)^2\frac{\mu}{L_{str}}+C^T\frac{\mu^2}{L_{str}^2}-C^T\left(B^T\right)^2-\frac{\mu}{L_{str}}\left(B^T\right)^2+\frac{\mu^3}{L_{str}^3}\right].
\end{split}
\end{equation}
 By substituting (\ref{eq::Cmu}) into the terms on the right-hand side of this equation, it is easy to see that each of these terms is smaller than the first term on the right-hand side. Therefore, we can apply perturbation theory to the (\ref{eq::Cconst}) constraint at the strong coupling scale and beyond.
The (\ref{eq::muconst}) constraint can be evaluated as follows:
\begin{equation}
    \frac{m^2\mu}{L^2}+g^2\left[\frac{\mu}{L^2}\left(\left(B^T\right)^2+B^T\phi\right)+\frac{\mu^2}{L^3}B^T+\frac{\mu^3}{L^4}\right]\sim\frac{g^2}{L}B^T\left(\left(B^T\right)^2+B^T\phi\right)
\end{equation}
Here we have replaced $C^T\sim B^T$. For scales $L>L_{str}$ the first term on the left-hand side dominates among all other terms on the same side. Once we reach the scales $L\sim L_{str}$, the nonlinear terms have the same order of magnitude as the linear term
\begin{equation}
    m^2\mu\sim g^2\mu \left(B^T\right)^2\sim g^2\frac{\mu^2}{L}B^T\sim g^2\frac{\mu^3}{L^2}
\end{equation}
and we can no longer perturbatively approximate the constraint. For $L<L_{str}$ the nonlinear terms dominate and we can estimate 
\begin{equation}
    \mu\sim \mathcal{O}(1)LB^T+\mathcal{O}(1)\phi+\mathcal{O}\left(\frac{1}{\sqrt{g}}\frac{L}{L_{str}}\right).
\end{equation}
Bearing in mind that the perturbation theory for the constraint (\ref{eq::Cconst}) holds for $L<L_{str}$, we obtain the Lagrangian density
\begin{equation}\label{eq::KBLang2}
    \mathcal{L}=\mathcal{L}_0+\mathcal{L}_{int}, \qquad\text{where}
\end{equation}
\begin{equation*}
    \begin{split}
        \mathcal{L}_0&=-\frac{1}{2}B_i^T(\Box+m^2)\frac{m^2}{-\Delta+m^2}B_i^T-\frac{1}{2}\phi(\Box+m^2)(-\Delta\phi)\qquad\text{and}\\\\
        \mathcal{L}_{int}&\sim g^2\left(B^T\right)^4+g^2\left(B^T\right)^3\frac{\phi}{L}+\mathcal{O}\left(g^2\left(B^T\right)^2\frac{\phi^2}{L^2}\right).
    \end{split}
\end{equation*}
 Note that both terms in $\mathcal{L}_{int}$ involve temporal and spatial derivatives in a manner similar to (\ref{eq::KBLang}). However, now that we have an additional contribution from the $\mu$ constraint, here we only present the power of the fields divided by the appropriate length scale. The first term provides us with the canonical normalisation for the transverse modes. Thus, the new canonically normalised variable is given by 
\begin{equation}
    B_n^T\sim gLB^T 
\end{equation}
and therefore, the minimal level of quantum fluctuations for the original field is given by 
\begin{equation}
    \delta_LB^T\sim\frac{1}{\sqrt{g}L}
\end{equation}
for scales $L<L_{str}$. 
Using these results, we can evaluate the corrections for the longitudinal mode below the strong coupling scale. These are given by the second term of $\mathcal{L}_{int}$, and can now be estimated as
\begin{equation}
    g^2\left(B^T\right)^3\frac{\phi}{L}\sim\frac{\sqrt{g}}{L^4}.
\end{equation}
It may be surprising that the corrections due to the strongly coupled mode remain outside the perturbation theory. The reason for this lies in the nature of the self-interaction we have considered. It is not gauge invariant, and hence if we were to compare the theories we have obtained with the massless case, we would find that the corrections remain. However, once the strongly coupled degrees of freedom enter the strong coupling regime, they lose their linear propagator, and stop evolving. This is analogous to the Vainshtein mechanism in massive gravity, where the longitudinal mode does not evolve beyond the strong coupling scale. In the next chapter, we will summarise and discuss the key findings and implications of the study.

\section{Discussion}
The aim of this work was to study the strong coupling of the interacting massive Kalb-Ramond and Proca theories. We saw that a similar mechanism applies to both theories when the massless limit is taken. 
In the framework of perturbation theory, the theories diverge in the massless limit due to the interactions. Since these theories were modified by the same type of interactions, divergences appear in the same powers of mass. Once the non-linear terms become of the order of the linear terms on the level of equations of motion, the modes which were the source of the discontinuity enter the strong coupling regime. Because of the same form of the interaction, the Vainshtein scale to which this corresponds is the same for both theories and is given by $L_{str}\sim\frac{\sqrt{g}}{m}$.  
 It is worth noting that a similar type of Vainshtein mechanism applies to massive gauge theories in general, such as the massive gravity and the massive Yang-Mills theory, where the mass is added by hand. 
The reason for the comparison between the two theories lies in the assertion of duality, which states that both theories describe the same physics. Given that the duality exists, one would expect the same number of degrees of freedom to behave identically. However, we have found that the opposite is true.
Due to the minimal level of quantum fluctuations, the discontinuity and the strong coupling do not appear for the same kind of modes. In the Kalb-Ramond theory, the transverse modes are the source of the discontinuity in the perturbative regime. As soon as they enter the strong coupling regime, the theory propagates only the longitudinal modes. For them the perturbation theory is valid at all scales. In the Proca theory, on the other hand, the discontinuity appears due to the longitudinal modes. They cross the scale of strong coupling and stop to propagate thereafter. Therefore, for scales smaller than the Vainshtein scale, Proca theory only propagates two transverse modes. Assuming that the coupling constant of the theory is smaller than unity, these modes never enter the strong coupling regime.
For this reason, there is a large discrepancy between these theories also for a non-zero mass, whose origin lies in the minimal level of quantum fluctuations for the same type of modes. 
This indicates that the duality between Proca and massive Kalb-Ramond theory is not present, contrary to the numerous claims in the literature. Given that this result was obtained in the context of self-interactions, the theories need to be further explored in a more general framework, which will be carried out in future work.  
\\\\
It could be interesting to see whether the duality persists for more general $p$-forms in $D$-dimensions. In the appendix we will present the results for massive three-forms. We have also seen in this paper how the Vainshtein mechanism is applied to the Kalb-Ramond field. Since this field naturally occurs in the context of massive Hermitian gravity, it would be interesting to see how its behavior changes when gravity and non-trivial backgrounds are taken into account.

\subsection*{Acknowledgements}
\textit{I am grateful to my supervisor Slava Mukhanov for pointing out to me this duality confusion, and for providing me with useful discussions during the work on this paper and comments on the draft of the paper. I would also like to thank Ali H. Chamseddine for sharing with me useful literature on the connection between the Kalb-Ramond field and gravity, Thomas Steingasser, for useful comments on the draft of paper and Alexander Vikman for useful discussions. This work was partially supported by the Deutsche Forschungsgemeinschaft (DFG, German Research Foundation) under Germany’s Excellence Strategy – EXC-2111 – 390814868.}

\section{Appendix: The duality of the massive 3-form}
It is also frequently claimed that there exists a duality between a massive 3-form and a scalar field \cite{Kawai, Trugenberger, Quevedo, 2001Smailagic, 2002Casini, Auria, Buchbinder, Dalmazi, Shifman, Garcia, Kuzenko, CF1980, C2019}. The action for the 3-form field is given by
\begin{equation}
    S=\int d^4x\left(-\frac{1}{48}W_{\mu\nu\alpha\beta}W^{\mu\nu\alpha\beta}+\frac{m^2}{12}C_{\mu\nu\alpha}C^{\mu\nu\alpha}\right)
\end{equation}
where $W_{\mu\nu\alpha\beta}=C_{\nu\alpha\beta,\mu}-C_{\mu\alpha\beta,\nu}+C_{\beta\mu\nu,\alpha}-C_{\alpha\mu\nu,\beta}$ is the field strength tensor and $C_{\mu\nu\alpha}$ is totally anti-symmetric. Starting from the parent action of the three-forms, the dualization procedure is used to obtain the action for the scalar field:
\begin{equation}\label{eq::scalaraction}
    S=\frac{1}{2}\int d^4x\left(\phi_{,\mu}\phi^{,\mu}-m^2\phi^2\right)
\end{equation}
Following the same way as before, we can see that the duality might not be present for these theories. In order to express the 3-form action in terms of physical modes, we will first separate the 3-form into a spatial part and a temporal part, with the spatial part being further decomposed as  
\begin{equation}
   \begin{split}
        &C_{0ij}=\varepsilon_{ijk}\left(C_k^T+\mu_{,k}\right), \qquad C_{k,k}^T=0 \qquad\text{and}\\\\
        &C_{ijk}=\varepsilon_{ijk}\chi
   \end{split}
\end{equation}
and arrive at the following Lagrangian density
\begin{equation}\label{eq::3formL1}
    \mathcal{L}=\frac{1}{2}\left[m^2C_i^TC_i^T-\mu\left(-\Delta+m^2\right)\Delta\mu-2\dot{\chi}\Delta\mu+\dot{\chi}\dot{\chi}-m^2\chi^2\right].
\end{equation}
Integrating out the non-propagating degrees of freedom, $\mu$ and $C_i^T$, (\ref{eq::3formL1}) becomes
\begin{equation}
    \mathcal{L}=-\frac{1}{2}\chi\left(\Box+m^2\right)\frac{m^2}{-\Delta+m^2}\chi.
\end{equation}
We can notice that the difference between the two kinetic terms is that (\ref{eq::3formL1}) is not canonically normalised. The canonically normalised variable is
 \begin{equation}
     \chi_n=\sqrt{\frac{m^2}{-\Delta+m^2}}\chi.
 \end{equation}
 After canonical normalization of the fields the two kinetic terms look the same. However, the minimal level of quantum fluctuations for the original fields $\phi$ and $\chi$ differs for scales $\frac{1}{L^2}\gg m^2$. On one side we have for the scalar field 
 \begin{equation}
     \delta \phi_{L}\sim\frac{1}{L},
 \end{equation}
 while on the other 
 \begin{equation}\label{eq::chifluct}
     \delta \chi_{L}\sim\frac{1}{mL^2}.
\end{equation}
The mass dependence only appears for the pseudo-scalar of the 3-form. Similarly to the contrast between Kalb-Ramond and Proca fields, this will be the source of the distinction for the two interacting theories.
\\
Now we shall add a self-interaction to of each of the theories. The actions are given by
\begin{equation}\label{eq::scalaraction2}
    S=\frac{1}{2}\int d^4x\left(\phi_{,\mu}\phi^{,\mu}-m^2\phi^2-\frac{g^2}{2}\phi^4\right)
\end{equation}
and
\begin{equation}\label{eq::3formL2}
    S=\frac{1}{12}\int d^4x\left[-\frac{1}{4}W_{\mu\nu\alpha\beta}W^{\mu\nu\alpha\beta}+m^2C_{\mu\nu\alpha}C^{\mu\nu\alpha}+\frac{g^2}{12}\left(C_{\mu\nu\alpha}C^{\mu\nu\alpha}\right)^2\right].
\end{equation}
To begin with, we will express the action \label{eq::3formL2} in terms of the propagating degree of freedom $\chi$ only. In a manner similar to the Kalb-Ramond field, there are two constraints which we obtain by the variation of the action with respect to $\mu$ and $C^T_i$:
\begin{equation}
    \begin{split}
        \left(-\Delta+m^2\right)\left(-\Delta\mu\right)=&\Delta\dot{\chi}+g^2\partial_i\left[\mu_{,i}\mu_{,j}\mu_{,j}+2\mu_{,i}\mu_{,j}C_j^T+C_i^T\mu_{,j}\mu_{,j}\right.\\\\
        &\left.+2C_i^TC_j^T\mu_{,j}+\left(\mu_{,i}+C_i^T\right)\left(C_j^TC_j^T-\chi^2\right)\right],
    \end{split}
\end{equation}

\begin{equation}
  \begin{split}
        m^2C_k^T=&-g^2P^T_{ik}\left[C_i^TC_j^TC_j^T+2C_i^TC_j^T\mu_{,j}+C_j^TC_j^T\mu_{,i}+2C_j^T\mu_{,j}\mu_{,i}\right.\\\\&\left.+\left(C_i^T+\mu_{,i}\right)\left(\mu_{,j}\mu_{,j}-\chi^2\right)\right].
  \end{split}
\end{equation}
By perturbatively resolving these conditions and inserting the solution into (\ref{eq::3formL2}), we obtain the Lagrangian density
\begin{equation}\label{eq::3formL3}
    \mathcal{L}=\mathcal{L}_0+\mathcal{L}_{int},\qquad\text{where}
\end{equation}
\begin{equation*}
   \begin{split}
        \mathcal{L}_0&=-\frac{1}{2}\chi\left(\Box+m^2\right)\frac{m^2}{-\Delta+m^2}\chi \qquad\text{and}\\\\
        \mathcal{L}_{int}&=\frac{g^2}{4}\left[D\left(\dot{\chi}_{,i}\right)D\left(\dot{\chi}_{,i}\right)\right]^2-\frac{g^2}{2}D\left(\dot{\chi}_{,i}\right)D\left(\dot{\chi}_{,i}\right)\chi^2+\frac{g^2}{4}\chi^4+\mathcal{O}\left(\frac{g^4\chi^6}{m^2}\right),
   \end{split}
\end{equation*}
where the operator $D$ is given by (\ref{eq::Dop}). Here we have only kept the most dominant terms. Comparing the interacting terms with the kinetic ones on scales $\frac{1}{L^2}\gg m^2$  and considering (\ref{eq::chifluct}), we find the strong coupling scale of the mode $\chi$ which is given by 
\begin{equation}
    L_{str}\sim\frac{\sqrt{g}}{m}.
\end{equation}
At this scale, also the nonlinear terms of the form $g^{2n}\frac{\chi^{2n+2}}{m^{2n-2}}$ with $n=2,3,...$ become dominant. This is due to the $C^T$ constraint for which the perturbation expansion is no longer valid at the strong coupling scale. Solving the constraints in a similar way as in the former case of Kalb-Ramond field, the most dominant terms in the Lagrangian yielding the new canonical normalization for the 3-form scalar are of the form 
\begin{equation}
    \mathcal{L}_{int}\sim g^2\chi^4.
\end{equation}
It should be noted that this expression represents only the dependence on the longitudinal mode with the derivatives of it omitted. While the derivatives act on $\chi$, they cancel with the operator $D$ because of the dimensions. 
For scales below the strong coupling scale, the longitudinal mode will be in the strong coupling regime, with the minimal level of quantum fluctuations given by $\delta \chi_L\sim\frac{1}{\sqrt{g}L}$, similarly to the transverse modes of the massive Kalb-Ramond field. On the other hand, the scalar field $\phi$ does not enter the strong coupling region as long as $g\ll 1$. The reason for this difference lies in the minimal quantum fluctuations of the original fields. Thus, it is not possible to conclude that these theories are dual on the basis of the previous arguments.

\end{document}